\documentclass[12pt]{article}
\usepackage{a4wide}
\usepackage{graphicx}
\begin{document}
{\renewcommand{\thefootnote}{\fnsymbol{footnote}}
\begin{center}
{\LARGE  Symmetries of Space-time }\\
\vspace{1.5em}
Martin Bojowald\footnote{e-mail address: {\tt bojowald@gravity.psu.edu}}
\\
\vspace{0.5em}
Institute for Gravitation and the Cosmos,\\
The Pennsylvania State
University,\\
104 Davey Lab, University Park, PA 16802, USA\\
\vspace{1.5em}
\end{center}
}

\setcounter{footnote}{0}

\begin{abstract}
  The equations of Hamiltonian gravity are often considered ugly cousins of
  the elegant and manifestly covariant versions found in the Lagrangian
  theory. However, both formulations are fundamental in their own rights
  because they make different statements about the nature of space-time and
  its symmetries. These implications, along with the history of their
  derivation and an introduction of recent mathematical support, are discussed
  here.
\end{abstract}

General relativity is distinguished by its covariance under space-time
diffeomorphisms, a large set of symmetries which guarantees coordinate
independence and supplies fruitful links between physics and geometry.
However, the symmetries are different in the Lagrangian and Hamiltonian
pictures. Throughout an interesting history of work on Hamiltonian gravity,
this under-appreciated state of affairs has led to pronouncements that verge
on the heretical. Dirac, for instance --- one of the outstanding protagonists
--- accompanied his detailed analysis in \cite{DiracHamGR} by ``It would be
permissible to look upon the Hamiltonian form as the fundamental one, and
there would then be no fundamental four-dimensional symmetry in the theory.''
He did not elaborate on this conclusion, but recent work in mathematics and
physics provides an updated picture. If we put together contributions by
relativists and mathematicians --- some older and some recent --- we can
confirm the prescient nature of Dirac's insights. At the same time, we improve
our fundamental understanding of space-time.

The history of Hamiltonian gravity had begun well before Dirac's entry,
spawned by questions about the analysis of the electromagnetic field. Starting
in 1929, Heisenberg and Pauli \cite{HeisenbergPauli,HeisenbergPauliII} had
applied canonical quantization to Maxwell's theory. An important issue was the
covariance of their formulation, as it still is in the case of
gravity. Rosenfeld \cite{Rosenfeld} presented a detailed analysis of
Hamiltonian general relativity, including a discussion of the important role
of constraints. After a gap of almost 20 years, Bergmann and his collaborators
turned the analysis of constraints into a program
\cite{NonLinFields,NonLinFieldsII,HamGREM,AndersonBergmann}, in parallel with
Dirac \cite{GenHamDyn1} not only in the timing of important work (1950) but
also in apparent heresies: according to \cite{NonLinFields} ``there is
probably no particular reason why the theory of relativity must appear in the
form of Riemannian geometry.'' The analysis of constraints most widely used
today was developed by Dirac, and applied by him to gravity
\cite{DiracHamGR}. Dirac was able to bring Rosenfeld's results to a more
convenient form by replacing general tetrads with metric variables adapted to
a spatial foliation. The final step was made by Arnowitt, Deser and Misner in
the 1960s \cite{ADM}, introducing a powerful parameterization of the
space-time metric by lapse $N$, shift $M^a$ and the metric $q_{ab}$ on a
spatial hypersurface. The resulting ADM formulation is widely used in
numerical relativity, cosmology, and quantum gravity.

An important question for Rosenfeld, following Heisenberg and Pauli, was the
role of symmetries. He was able to show that covariance implies constraints on
the fields, which are equivalent to some components of Einstein's
equation. However, he did not encounter the characteristic symmetry of
Hamiltonian gravity because his variables were not adapted to a space-time
foliation. Dirac was the first to introduce this crucial condition and to
derive the symmetries. In modern ADM notation, there are infinitely many
generators $G_{N,M^a}$, subject to commutator relations
\begin{equation}
 [G_{N_1,M_1^a},G_{N_2,M_2^a}]=G_{{\cal L}_{M_1}N_2-{\cal L}_{M_2}N_1,
 [M_1,M_2]^a+q^{ab}(N_1\partial_bN_2-N_2\partial_bN_1)}\,. \label{HH}
\end{equation}
While the first few terms show the typical form of Lie derivatives as
infinitesimal spatial diffeomorphisms, the last term is fundamentally
different. In particular, it contains the inverse spatial metric $q^{ab}$,
which is not a structure constant and not one of the generators. A
satisfactory mathematical formulation requires some care. It was provided only
recently \cite{ConsAlgebroid}, concluding that the brackets (\ref{HH}) belong
to a Lie algebroid.

In physics terminology, the relations (\ref{HH}) have ``structure functions''
depending on $q_{ab}$. As realized by Hojman, Kucha\v{r} and Teitelboim
\cite{Regained}, they present a new symmetry deforming spatial hypersurfaces,
tangentially (along $M^a$) and normally (along $Nn^{\mu}$, with the unit
normal $n^{\mu}$).  The symmetry agrees with space-time diffeomorphisms ``on
shell'' when equations of motion hold.  However, it is not identical with
space-time diffeomorphisms. Off-shell properties are relevant when we talk
about the Riemannian structure underlying general relativity, or the
4-dimensional symmetries of space-time. Is the symmetry generated by
(\ref{HH}) more fundamental, vindicating Dirac's heresy?  Or does it lead to
departures from Riemannian structures, justifying Bergmann's iconoclasticism?
Unfortunately, the importance of the new symmetry is often obscured by the
messy derivation of its relations (\ref{HH}).  Dirac first found them by
brute-force computations of Poisson brackets.  Kucha\v{r}
\cite{KucharHypI,KucharHypII,KucharHypIII} rederived them in terms of
commutators of derivatives by the functions that embed a spatial hypersurface
in space-time. Such derivations are long and do not easily suggest intuitive
pictures.

More recently, in 2010, a new derivation has been given by Blohmann, Barbosa
Fernandes, and Weinstein \cite{ConsAlgebroid}. Even though it derives a
central statement of Hamiltonian gravity, their method does not require an
explicit implementation of the $3+1$ split which often hides the elegance of
covariant theories. As presented in \cite{ConsAlgebroid}, spread over several
proofs of other results, the new derivation is not easy to access. The
following two paragraphs present a remodeled version in compact form, painted
in notation cherished by relativists.

Choose a Riemannian space-time with signature $\epsilon=\pm 1$, pick a spatial
foliation, and introduce Gaussian coordinates adapted to one of the spatial
hypersurfaces. The resulting line element ${\rm d}s^2=\epsilon{\rm
  d}t^2+q_{ab}{\rm d}x^a{\rm d}x^b$ depends only on the spatial metric
$q_{ab}$. Its general form is preserved by any vector field $v^{\rho}$ which
satisfies $n^{\mu}{\cal L}_vg_{\mu\nu}=0$, using the unit normal
$n_{\mu}=({\rm d}t)_{\mu}$ in the Gaussian system.  We expand this condition
by writing out the Lie derivative:
\begin{equation}
 0=n^{\mu}{\cal L}_vg_{\mu\nu}=n^{\mu}v^{\rho}\partial_{\rho}g_{\mu\nu}+
 n^{\mu}g_{\nu\rho}\partial_{\mu}v^{\rho}+n^{\mu}g_{\mu\rho}\partial_{\nu}v^{\rho}\,.
\end{equation}
In the first term, we use
$n^{\mu}v^{\rho}\partial_{\rho}g_{\mu\nu}=v^{\rho}\partial_{\rho}n_{\nu}-g_{\mu\nu}v^{\rho}\partial_{\rho}n^{\mu}$
and manipulate the last term to
$n^{\mu}g_{\mu\rho}\partial_{\nu}v^{\rho}= \partial_{\nu}(n^{\mu}v^{\rho}g_{\mu\rho})-v^{\rho}\partial_{\nu}n_{\rho}$.
Combining these equations and using ${\rm d}n_{\mu}=({\rm d}^2t)_{\mu}=0$, we
arrive at
\begin{equation} \label{Gauss}
 0=n^{\mu}{\cal L}_vg_{\mu\nu}=[n,v]^{\mu}g_{\mu\nu}+\partial_{\nu}(n^{\mu}v^{\rho}g_{\mu\rho})\,.
\end{equation}

We now decompose $v^{\mu}=Nn^{\mu}+M^{\mu}$ into components normal and
tangential to the foliation. (We have $M^{\mu}=M^as_a^{\mu}$ if $s_a^{\mu}$,
$a=1,2,3$, is a spatial basis.)  Equation (\ref{Gauss}) then implies
$n^{\mu}\partial_{\mu}N=0$ and $[n,M]^{\mu}=-\epsilon
q^{\mu\nu}\partial_{\nu}N$.  These new equations, together with linearity and
the Leibniz rule, allow us to write the Lie bracket of two vector fields,
$v_1^{\mu}=N_1n^{\mu}+M_1^{\mu}$ and $v_2^{\mu}=N_2n^{\mu}+M_2^{\mu}$, as
\begin{equation} \label{vv}
  [v_1,v_2]^{\mu}=\left({\cal L}_{M_1}N_2-
{\cal L}_{M_2}N_1\right)n^{\mu}+[M_1,M_2]^{\mu}-\epsilon q^{\mu\nu}(N_1\partial_{\nu}N_2-N_2\partial_{\nu}N_1)\,.
\end{equation}
The result agrees with (\ref{HH}) for $\epsilon=-1$, while $\epsilon=1$
corresponds to the version of (\ref{HH}) in Euclidean general relativity.

We are left with the problem of structure functions in the brackets. They are
not constant because the spatial metric changes under our symmetries.  If we
cannot fix $q_{ab}$, we have to deal with the abundance of infinitely many
copies of the brackets (\ref{vv}), one for each $q_{ab}$.  Our new-found
riches can be invested in a fancy mathematical structure: The brackets are
defined on sections of an infinite-dimensional vector bundle with fiber
$(N,M^a)$ and as base manifold the space of spatial metrics.

A heuristic argument shows that this viewpoint is fruitful: Assume finitely
many constraints $C_I$, $I=1,\ldots n$, defined on a phase space $B$, with
Poisson brackets $\{C_I,C_J\}=c_{IJ}^K(x)C_K$ for $x\in B$.  Extend the
generators by introducing, iteratively,
$C_{HIJ\cdots}:=\{C_H,C_{IJ\cdots}\}$. The new system has infinitely many
generators with structure {\em constants} because $\{C_I,C_J\}=C_{IJ}$ and so
on.  All these generators can be written as the original constraints
multiplied with functions on $B$. They are examples of a new kind of vector
field, or sections $\alpha=\alpha^IC_I$ of a vector bundle over $B$ with fiber
coordinates $\alpha^I(x)$.  There is a Lie bracket
$[\alpha_1,\alpha_2]=\{\alpha_1^IC_I,\alpha_2^JC_J\}$, and the linear map
$\rho_{\alpha}={\cal L}_{X_{\alpha^IC_I}}$ from $\alpha$ to the Lie derivative
along the Hamiltonian vector field of $\alpha^IC_I$ is a Lie-algebra
homomorphism. It cooperates with the bracket in a Leibniz rule:
$[\alpha_1,g\alpha_2]=g[\alpha_1,\alpha_2]+(\rho_{\alpha_1}g)\alpha_2$.  These
properties characterize the vector bundle as a Lie algebroid \cite{Pradines}.

The brackets of Hamiltonian gravity form a Lie algebroid.  It is the
infinitesimal version of the Lie groupoid of finite evolutions, pasting
together whole chunks of space-time between spatial hypersurfaces
\cite{ConsAlgebroid}. At this point, two important research directions are
merging, the physical analysis of Hamiltonian gravity and the mathematical
study of Lie algebroids. The link remains rather unexplored, but it shows
great promise. And it could help us to illuminate Dirac's statement.

As for the promise, a good understanding of the right form of Lie algebroid
representations could show the way to a consistent theory of canonical quantum
gravity. It is already clear that there is fascinating physics behind the
math. Lie algebroids can be deformed more freely than Lie algebras. A
gravitational example is given by the relations (\ref{HH}), where a free
phase-space function $\beta$ multiplying $q^{ab}$ can be inserted.  We do not
always obtain new versions of space-time: generators can be redefined so as to
absorb $\beta$ \cite{Absorb}, but only if this function does not change sign
anywhere. If it does, for instance at large curvature in models of quantum
gravity \cite{JR,ScalarHol,HigherSpatial}, a smooth transition from
$\epsilon=-1$ to $\epsilon=1$ in (\ref{vv}) implies a passage from Lorentzian
space-time to Euclidean 4-space \cite{Action,SigChange,SigImpl}. Such a model
with non-singular signature change cannot be Riemannian. Bergmann's
expectation has been confirmed.

What about Dirac's heresy? Is the Hamiltonian form more fundamental than the
Lagrangian one? It is hard to realize space-time structures with
$\beta$-modified brackets in Lagrangian form: An action principle needs a
measure factor, such as ${\rm d}^4x \sqrt{|\det g|}$, but a non-Riemannian
version corresponding to brackets with $\beta\not=1$ remains unknown. The
Hamiltonian version has no such problems, and may well be considered more
fundamental.

But is it realized in nature? Only a consistent version of canonical quantum
gravity can give a final answer.

\section*{Acknowledgements}

This work was supported in part by NSF grant PHY-1307408.


\begin{thebibliography}{10}

\bibitem{DiracHamGR}
P.~A.~M.\ Dirac,
\newblock The theory of gravitation in Hamiltonian form,
\newblock {\em Proc.\ Roy.\ Soc.\ A} 246 (1958) 333--343

\bibitem{HeisenbergPauli}
W.\ Heisenberg and W.\ Pauli,
\newblock Zur Quantendynamik der Wellenfelder,
\newblock {\em Z.\ Phys.} 56 (1929) 1--61

\bibitem{HeisenbergPauliII}
W.\ Heisenberg and W.\ Pauli,
\newblock Zur Quantendynamik der Wellenfelder II,
\newblock {\em Z.\ Phys.} 59 (1930) 168--190

\bibitem{Rosenfeld}
L.\ Rosenfeld,
\newblock Zur Quantelung der Wellenfelder,
\newblock {\em Annalen Phys.} 5 (1930) 113--152

\bibitem{NonLinFields}
P.~G.\ Bergmann,
\newblock Non-Linear Field Theories,
\newblock {\em Phys.\ Rev.} 75 (1949) 680--685

\bibitem{NonLinFieldsII}
P.~G.\ Bergmann and J.~H.~M.\ Brunings,
\newblock Non-Linear Field Theories II. Canonical Equations and Quantization,
\newblock {\em Rev.\ Mod.\ Phys.} 21 (1949) 480--487

\bibitem{HamGREM}
P.~G.\ Bergmann, R.\ Penfield, R.\ Schiller, and H.\ Zatzkis,
\newblock The Hamiltonian of the General Theory of Relativity with
  Electromagnetic Field,
\newblock {\em Phys.\ Rev.} 80 (1950) 81--88

\bibitem{AndersonBergmann}
J.~L.\ Anderson and P.~G.\ Bergmann,
\newblock Constraints in Covariant Field Theories,
\newblock {\em Phys.\ Rev.} 83 (1951) 1018--1025

\bibitem{GenHamDyn1}
P.~A.~M.\ Dirac,
\newblock Generalized Hamiltonian dynamics,
\newblock {\em Can.\ J.\ Math.} 2 (1950) 129--148

\bibitem{ADM}
R.\ Arnowitt, S.\ Deser, and C.~W.\ Misner,
\newblock The Dynamics of General Relativity, In L.\ Witten, editor, {\em
  Gravitation: An Introduction to Current Research},
\newblock Wiley, New York, 1962,
\newblock Reprinted in \cite{ADMRe}

\bibitem{ConsAlgebroid}
C.\ Blohmann, M.~C.\ Barbosa~Fernandes, and A.\ Weinstein,
\newblock Groupoid symmetry and constraints in general relativity. 1:
  kinematics,
\newblock {\em Commun.\ Contemp.\ Math.} 15 (2013) 1250061, [arXiv:1003.2857]

\bibitem{Regained}
S.~A.\ Hojman, K.\ Kucha\v{r}, and C.\ Teitelboim,
\newblock Geometrodynamics Regained,
\newblock {\em Ann.\ Phys.\ (New York)} 96 (1976) 88--135

\bibitem{KucharHypI}
K.~V.\ Kucha\v{r},
\newblock Geometry of hypersurfaces. I,
\newblock {\em J.\ Math.\ Phys.} 17 (1976) 777--791

\bibitem{KucharHypII}
K.~V.\ Kucha\v{r},
\newblock Kinematics of tensor fields in hyperspace. II,
\newblock {\em J.\ Math.\ Phys.} 17 (1976) 792--800

\bibitem{KucharHypIII}
K.~V.\ Kucha\v{r},
\newblock Dynamics of tensor fields in hyperspace. III,
\newblock {\em J.\ Math.\ Phys.} 17 (1976) 801--820

\bibitem{Pradines}
J.\ Pradines,
\newblock Th\'eorie de Lie pour les groupo\"{\i}des diff\'erentiables. Calcul
  diff\'erenetiel dans la cat\'egorie des groupo\"{\i}des infinit\'esimaux,
\newblock {\em Comptes Rendus Acad. Sci. Paris S\'er.\ A--B} 264 (1967)
  A245--A248

\bibitem{Absorb}
R.\ Tibrewala,
\newblock Inhomogeneities, loop quantum gravity corrections, constraint algebra
  and general covariance,
\newblock {\em Class.\ Quantum Grav.} 31 (2014) 055010, [arXiv:1311.1297]

\bibitem{JR}
J.~D.\ Reyes,
\newblock {\em Spherically Symmetric Loop Quantum Gravity: Connections to
  2-Dimensional Models and Applications to Gravitational Collapse},
\newblock PhD thesis, The Pennsylvania State University, 2009

\bibitem{ScalarHol}
T.\ Cailleteau, J.\ Mielczarek, A.\ Barrau, and J.\ Grain,
\newblock Anomaly-free scalar perturbations with holonomy corrections in loop
  quantum cosmology,
\newblock {\em Class.\ Quant.\ Grav.} 29 (2012) 095010, [arXiv:1111.3535]

\bibitem{HigherSpatial}
M.\ Bojowald, G.~M.\ Paily, and J.~D.\ Reyes,
\newblock Discreteness corrections and higher spatial derivatives in effective
  canonical quantum gravity,
\newblock {\em Phys.\ Rev.\ D} 90 (2014) 025025, [arXiv:1402.5130]

\bibitem{Action}
M.\ Bojowald and G.~M.\ Paily,
\newblock Deformed General Relativity and Effective Actions from Loop Quantum
  Gravity,
\newblock {\em Phys.\ Rev.\ D} 86 (2012) 104018, [arXiv:1112.1899]

\bibitem{SigChange}
J.\ Mielczarek,
\newblock Signature change in loop quantum cosmology,
\newblock {\em Springer Proc.\ Phys.} 157 (2014) 555, [arXiv:1207.4657]

\bibitem{SigImpl}
M.\ Bojowald and J.\ Mielczarek,
\newblock Some implications of signature-change in cosmological models of loop
  quantum gravity,
\newblock {\em JCAP} 08 (2015) 052, [arXiv:1503.09154]

\bibitem{ADMRe}
R.\ Arnowitt, S.\ Deser, and C.~W.\ Misner,
\newblock The Dynamics of General Relativity,
\newblock {\em Gen.\ Rel.\ Grav.} 40 (2008) 1997--2027

\end{thebibliography}
\end{document}